\documentclass[preprint,12pt]{elsarticle}
\usepackage{amssymb}
\usepackage{amsmath}
\usepackage{amsfonts}
\usepackage{romannum}
\usepackage{titlesec}
\usepackage{enumitem}
\titleformat{\section}{\normalfont\bfseries\scshape}{\thesection}{1em}{}
\usepackage{fancyhdr}

\begin{document}

\begin{frontmatter}

\title{Gauge invariance for a class of tree diagrams in the standard model}

\author[a]{Tai Tsun Wu\corref{correspondingauthor}}
\author[b]{Sau Lan Wu}

\cortext[correspondingauthor]{Corresponding author. \\ E-Mail address: tai.tsun.wu@cern.ch}

\address[a]{Gordon McKay Laboratory, Harvard University, \\Cambridge, MA~02138, U.~S.~A.}

\address[b]{Physics Department, University of Wisconsin-Madison,\\ Madison, WI~53706, U.~S.~A.}

\pagenumbering{arabic}

\begin{abstract}
For gauge theory, the matrix element for any physical process is independent of the gauge used.  Since this is a formal statement and examples are known where gauge invariance is violated, for any specific process this gauge invariance needs to be checked by explicit calculation.  In this paper, gauge invariance is found to hold for a large non-trivial class of processes described by tree diagrams in the standard model -- tree diagrams with two external $W$ bosons and any number of external Higgs bosons.  This verification of gauge invariance is quite complicated, and is based on a direct study of the difference between different gauges through induction on the number of external Higgs bosons.
\end{abstract}

\end{frontmatter}


\section{Introduction}
\thispagestyle{fancy}
\fancyhf{}
\rhead{CERN-TH-2018-198 \\}

In a recent paper~\cite{ttwu201}, the decay process 
\begin{equation}
\begin{aligned}
H \rightarrow \gamma\gamma
\label{aba:eq1.1}
\end{aligned}
\end{equation}
\noindent through one $W$ loop was studied in detail in the standard model of Glashow, Weinberg, and Salam~\cite{sglashow}. Here $H$ is the Higgs particle proposed theoretically in 1964~\cite{FERB} and discovered experimentally in 2012~\cite{atlasdisc}. This reference~\cite{ttwu201} gives the first case where the non-zero difference is found between the matrix element calculated using the $R_\xi$ gauge and the unitary gauge. This calculation, using Feynman rules for both gauges, is straightforward and elementary, involving only high-school algebra. 

In spite of its simplicity, this result raises immediately a number of questions, including the following: 

(1) What other matrix elements in the standard model have the property that the result using the $R_\xi$ gauge differs from that using the unitary gauge?

(2) The straightforward and elementary calculation presented in Ref. ~\cite{ttwu201} has the shortcoming that it gives no indication as to what the underlying reason is for the difference between the unitary gauge and the $R_\xi$ gauge. Is there another method of calculation that may be more enlightening?

For the first question here, the answer is of course there are many other matrix elements with this property. A second example has been given in Ref. ~\cite{ttwu201}, that for the decay 
\begin{equation}
\begin{aligned}
H \rightarrow Z\gamma,
\label{aba:eq1.2}
\end{aligned}
\end{equation}

\noindent also through a $W$ loop. Moreover any matrix element that contains an $H\gamma\gamma$ or an $H Z \gamma$ one-$W$-loop insertion has this property. But it is believed that there are many other such cases. Some possible candidates are the one $W$-loop matrix element for the processes 

\begin{equation}
\begin{aligned}
\gamma\gamma \rightarrow \gamma\gamma
\label{aba:eq1.3}
\end{aligned}
\end{equation}

\begin{equation}
\begin{aligned}
\gamma\gamma \rightarrow Z\gamma
\label{aba:eq1.4}
\end{aligned}
\end{equation}

\noindent etc. 

It is the purpose of the present paper to address the second question above. 

In the case of the decay  (\ref{aba:eq1.1}), the difference of the matrix element using the two gauges takes a very simple form -- see Eq.~(82) of Ref.~\cite{ttwu201}. Therefore, it seems reasonable to be able to find this difference directly without carrying out a tedious subtraction of matrix element in the two gauges. In order to learn a better way to find this difference, it is desirable and perhaps necessary to study simple cases first. What can be simple than the decay process (\ref{aba:eq1.1})? The answer is obvious: tree diagrams. 

It is therefore proposed in this paper to study an especially simple case of tree diagrams: those with two external $W$ boson and any number of external Higgs bosons. 

In Ref.~\cite{ttwu201}, the difference between the two gauges is attributed to the failure for a limiting process and a momentum integration to commute with each other. In the case of tree diagrams, there is no momentum integration and hence this lack of commutation is not relevant. Nevertheless, even without any integration, the present treatment of this particular class of tree diagrams is by no means straightforward. It is expected that the method developed here will be generalized to deal with diagrams with loops. 

In Sec. 2, the motivation and the basic idea of the present approach is described and motivated: it involves first the differentiation with respect to $\xi$, the parameter in the $R_\xi$ gauge, and secondly the fact that, for tree diagrams, the matrix elements are necessarily rational functions of $\xi$. In Sec. 3, this method is applied to this particular class of tree diagrams being studied. In this way, the derivative of the matrix elements is expressed in terms of two functions called $P$ and $Q$. Mathematical induction with respect to the number of external Higgs bosons is then applied to the difference $P-Q$. This rather involved induction process is described in the lengthy Sec. 4. Sec. 5 gives the conclusion and some discussions.

\section{Method}

For the purpose of the present study of comparing different gauges, what is the major difference between the $R_\xi$ and the unitary gauge? Since the unitary gauge is the limit $\xi \rightarrow \infty$ for the $R_\xi$ gauge, there can be no difference between these two gauges unless this variable appears in the matrix element for the $R_\xi$ gauge. 

For tree diagrams, the absence of loops implies that the Faddeev-Popov ghost ~\cite{lfadv} cannot appear. Without the Faddeev-Popov ghost, this variable $\xi$ appears only in the propagators for $W$ and $Z$ together with those for the associated 
$\varphi$ and $\varphi_0$, sometimes referred to as the Higgs ghosts. Therefore, for tree diagrams without any of these propagators, the same matrix element is obtained for the unitary gauge and the $R_\xi$ gauge. 

\begin{figure}
\begin{center}
\includegraphics[width=5in]{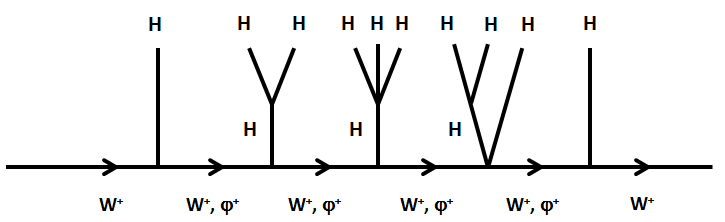}
\end{center}
\caption{Examples of $WWnH$ tree diagrams. Here n=10. }
\label{fig1}
\end{figure}

The simplest non-trivial cases are those with only a pair of external $W$ lines or a pair of external $Z$ lines without any additional vector-particle external lines. For definiteness, the case of two external $W$ lines is to be studied here. With this choice, there is a well-defined charged line that connects these two external $W$ lines; without loss of generality, this lines is taken to be positively charged, as illustrated in Fig.~\ref{fig1}. When the $R_\xi$ gauge is used, the propagators along this charged line is that of either $W^+$ or $\varphi^+$. 

For the tree diagrams in the standard model, this consideration leads to the $WWnH$ diagrams, meaning the tree diagrams with two external $W$ boson and any number of external Higgs boson. Such tree diagrams can look quite complicated, some examples being shown in Fig.~\ref{fig1}. It should be emphasized that both external $W$ bosons and all the external Higgs bosons are on their respective mass shells.  

In all the diagrams in this paper, including those in this Fig.~\ref{fig1}, the following convention is used unless otherwise specified: the horizontal line means a charged line, either $W$ or $\varphi$, with the positive charge going from left to right, while all lines not horizontal are Higgs lines. 

The only conceivable way to study these $WWnH$ matrix elements for a general value of $n$ is to carry out a mathematical induction on the number of $n$ of external Higgs bosons, i.e., to express the matrix element for any integer $n$ in terms of those with smaller values of $n$. However, as seen from the examples of Fig.~\ref{fig1}, the $W$ lines on the extreme right and on the extreme left are external and hence must be on mass shell, while those in between are not. Therefore, for the matrix elements themselves, there is no way to carry out a mathematical induction on this integer $n$. 

In order to apply induction to the present problem of $WWnH$, it is necessary to break the horizontal $W^+ / \phi^+$ line. For this purpose, it is proposed to make use of the following general property of tree diagrams. Since there is no momentum integration, the contribution of any tree diagram to the matrix element is a rational function of the parameter $\xi$. Moreover, since, for any fixed value of $n$, the matrix element for the $WWnH$ vertex is the sum of contributions from a finite number of tree diagrams, this matrix element is itself a rational function of $\xi$. 

A rational function of $\xi$ is either a constant or not a constant. If the matrix element for this $WWnH$ vertex is a constant, then this constant gives the answer for both the $R_\xi$ gauge and the unitary gauge, implying that there is no violation of gauge invariance. If there is a violation of  gauge invariance, i. e., if the matrix element calculated using the unitary gauge is different from that using the $R_\xi$ gauge, then the matrix element is not a constant. This not being a constant has the strong implication that not only the unitary gauge and the $R_\xi$ gauge give different answers, but also the answer obtained using the $R_\xi$ gauge must depend on the value of $\xi$.

It is in general easier to determine whether a complicated expression is zero or not then to find out whether it is a constant. For example, in the former problem, common factors can often be ignored. 

The central idea of the present investigation of the $WWnH$ vertices is therefore: apply the differential operator $\frac{\partial}{\partial\xi}$ to the matrix elements from tree diagrams. 

As to be shown in the present paper, this derivative with respect to $\xi$ of the tree matrix elements has a number of additional desirable properties. In particular, while it is not possible to apply induction with respect to $n$ directly to the matrix elements as discussed above, this induction procedure can be used on this derivative of the matrix element with respect to $\xi$.

While this idea of differentiating with respect to $\xi$ is motivated and justified only in the case of tree diagrams, it is likely that it is also useful in the cases of matrix elements given by diagrams with loops. 


\section{Differentiation of matrix elements with respect to $\xi$}

The first step is to apply this $\partial/\partial\xi$ to the $WWnH$ matrix elements so that an induction can be carried out on the number $n$ of external Higgs lines.

\begin{figure}
\begin{center}
\includegraphics[width=5in]{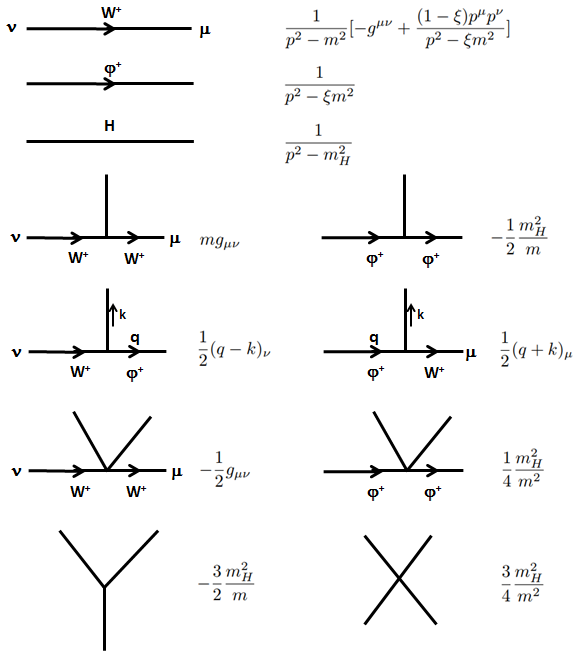}
\end{center}
\caption{Relevant Feynman rules for the $WWnH$ vertices in the $R_\xi$ gauge. }
\label{fig2}
\end{figure}

The relevant Feynman rules for the standard model~\cite{sglashow} are given in Fig.~\ref{fig2}. Following the notation of Ref.~\cite{ttwu201}, the coupling constant $g$ and the overall factors of $i$ have been omitted. Here $m$ is the mass of the $W$ boson, and $m_{H}$ is the mass of the Higgs boson. As seen from this Fig.~\ref{fig2}, the parameter $\xi$ appears only in the $W^+$ and $\varphi^+$ propagators, both in the denominator. Their derivatives with respect to $\xi$  are given by 

\begin{equation}
\begin{aligned}
\frac{\partial}{\partial\xi} \{ \frac{1}{p^2-m^2} [-g^{\mu\nu} + \frac{(1-\xi) p^\mu p^\nu}{p^2-\xi m^2}] \} = -\frac{1}{(p^2-\xi m^2)^2} p^\mu p^\nu
\label{aba:eq3.1}
\end{aligned}
\end{equation}

\noindent and

\begin{equation}
\begin{aligned}
\frac{\partial}{\partial\xi} \frac{1}{p^2-\xi m^2} = \frac{m^2}{(p^2-\xi m^2)^2}.
\label{aba:eq3.2}
\end{aligned}
\end{equation}

\noindent The similarity of the right-hand sides of these Eqs.~(\ref{aba:eq3.1}) and (\ref{aba:eq3.2}), both involving the factor $(p^2-\xi m^2)^2$ in the denominators, plays an important role in the present treatment. In particular, this appearance of the square indicates that the differentiation of the propagators with different momenta can be treated independently. 

Suppose a $\varphi^+$ propagator with momentum $p$ is differentiated with respect to $\xi$. This contribution to the derivative of the matrix element is the product of the following three factors: 

\begin{enumerate}[label=(\alph*)]
\item the right-hand side of Eq.~(\ref{aba:eq3.2});
\item a factor $Q$ coming from the part of the diagram to the right of this $\varphi^+$ propagator  with momentum $p$; and  
\item a factor $Q'$ coming from the part of the diagram to the left of this $\varphi^+$ propagator with  momentum $p$. 
\end{enumerate}


\noindent With the convention that the factor (a) is not written explicitly, this contribution to the derivative is $QQ'$. 

Consider next a $W^+$ propagator with the same momentum $p$. Let the Lorentz indices for this $W^+$ propagator be denoted by $\nu$ to the left and $\mu$ to the right; see Fig. 2. This contribution to the derivative of the matrix element also consists of three factors: 

\vspace{12pt}

 (a') ~the same as (a);
 \vspace{4mm}
 
 (b') ~a factor $P$ coming from $p^\mu/m$ contracted with the Feynman rules coming from the part of the diagram to the right of this $W^+$ propagator with momentum $p$; and  
 \vspace{4mm}
 
 (c') ~a factor $P'$ coming from a similar contraction of $p^\nu/m$ with the corresponding left part of the diagram. 
 
 \noindent The contribution to the derivative is $PP'$; the signs are arranged so that the contributions from differentiating the $W^+$ propagator and the $\varphi^+$ propagator taken together is $PP' - QQ' $.

These $P$ and $Q$ of course depend on the number of external Higgs bosons to the right of the propagator that has been differentiated, and similarly these $P'$ and $Q'$ on the number of external Higgs bosons to the left. Roughly speaking, through the use of $\partial/\partial\xi$, each contribution to the $WWnH$ matrix element is broken up into two pieces, a right-hand piece ($P$ or $Q$) and a left-hand piece ($P'$ or $Q'$).

Through a trivial application of the $CPT$ Theorem~\cite{schwinger} or by looking at the Feynman rules of Fig.~\ref{fig2}, the $P'$ and $Q'$ can be expressed in terms of some $P$ and $Q$ with different momenta for the external lines. Therefore it is sufficient to concentrate on the $P$ and $Q$.

For each $P$ and each $Q$, the horizontal line on the extreme right is necessarily a $W^+$ line, and this $W^+$ line represents an external outgoing $W^+$ which is on mass shell. This is entirely similar to the case of the matrix element as discussed in Sec. 2. However, the horizontal line on the extreme left does not represent an external line. Therefore, the $P's$ and $Q's$ can be reduced starting from the left-hand side. Such is the great power of differentiation with respect to $\xi$. 

This somewhat complicated reduction of the $P's$ and $Q's$ is to be discussed in the next section.

\section{Properties of $P's$ and $Q's$} 

\noindent{\bf A. Simplest case}

The simplest of the $P's$ and $Q's$ are those that correspond to $n=1$. There is only one of each, they are conveniently designated as $P^{(1)}$, and $Q^{(1)}$, and their diagrammatic representations are shown in Fig. 3. For $P^{(1)}$, the left-hand line is a $W^+$, and a dot near this line is introduced to indicate the factor $p_2^\beta/m$; see (b') of Sec. 3. This dot also serves the purpose of showing that the diagram is for $P$, not for the matrix element itself. Thus a dot is also introduced for $Q$; for $Q^{(1)}$, the left-hand line is a $\varphi^+$, but the dot does not give any additional factor. 

From the Feynman rules of Fig. 2, it is seen immediately that 
\begin{equation}
\begin{aligned}
P^{(1)} = \frac{p_2^\beta}{m} m g_{\beta\alpha} = p_{2\alpha}
\label{aba:eq4.1}
\end{aligned}
\end{equation}

\noindent and

\begin{equation}
\begin{aligned}
Q^{(1)} = \frac{1}{2} (p_2+k)_\alpha = p_{2\alpha}
\label{aba:eq4.2}
\end{aligned}
\end{equation}

\begin{figure}
\begin{center}
\includegraphics[width=5in]{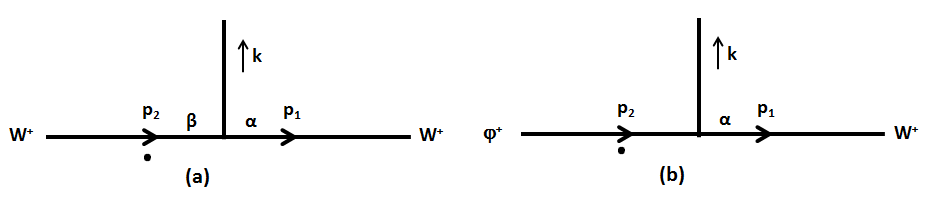}
\end{center}
\caption{Diagrams for (a) $P^{(1)}$ and (b) $Q^{(1)}$. }
\label{fig3}
\end{figure}

\noindent because $p_1$ is on mass shell, which implies

\begin{equation}
\begin{aligned}
p_{1\alpha} = 0.
\label{aba:eq4.3}
\end{aligned}
\end{equation}

\noindent The important relation

\begin{equation}
\begin{aligned}
P^{(1)} - Q^{(1)} = 0
\label{aba:eq4.4}
\end{aligned}
\end{equation}

\noindent then follows. It should be emphasized that this Eq.~(\ref{aba:eq4.4}) holds even for the external Higgs line off-shell. 

For $n=2$, there are four $P_j^{(2)}$ and $Q_j^{(2)}$, not counting the permutation of the external Higgs lines. The diagrams for these $P_j^{(2)}$ and $Q_j^{(2)}$ are given in Fig. 4. The total contributions for $n=2$ are 

\begin{equation}
\begin{aligned}
P^{(2)} = P^{(2)}_1 + \hat{P}^{(2)}_1 + P^{(2)}_2 + \hat{P}^{(2)}_2 + P^{(2)}_3 + P^{(2)}_4
\label{aba:eq4.5}
\end{aligned}
\end{equation}

\noindent and

\begin{equation}
\begin{aligned}
Q^{(2)} = Q^{(2)}_1 + \hat{Q}^{(2)}_1 + Q^{(2)}_2 + \hat{Q}^{(2)}_2 + Q^{(2)}_3,
\label{aba:eq4.6}
\end{aligned}
\end{equation}

\noindent where the $\hat{}$ denotes exchange of the two external Higgs bosons. 

The numbers of these $P_j^{\prime}s$ and $Q_j^{\prime}s$ increase rapidly with $n: 13~ P_j^{(3)}$ and $12~ Q_j^{(3)}$; $48~ P_j^{(4)}$ and $45~ Q_j^{(4)}$. All these $P_j^{\prime}s$ and $Q_j^{\prime}s$ up to $n=4$ have been studied in detail. 

\begin{figure}
\begin{center}
\includegraphics[width=5in]{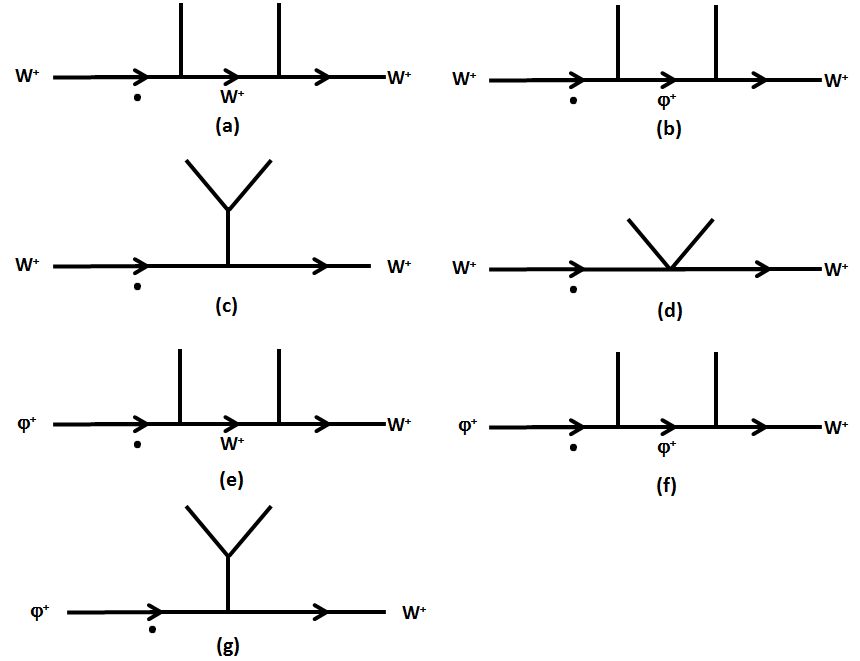}
\end{center}
\caption{Diagrams for (a) $P^{(2)}_1$, (b) $P^{(2)}_2$, (c) $P^{(2)}_3$, (d) $P^{(2)}_4$, (e) $Q^{(2)}_1$, (f) $Q^{(2)}_2$, and (g) $Q^{(2)}_3$. }
\label{fig4}
\end{figure}

\vspace{6pt}

\noindent {\bf B. Branches of the tree}

It is seen clearly from Fig. 1 that the tree diagram has many "branches", and these branches are formed entirely of the Higgs propagators and Higgs external lines. Some examples of these branches, taken mostly from Fig. 1, are shown in Fig. 5. These branches, and many others, can occur not only with the matrix elements, but also with the $P^{\prime}s$ and $Q^{\prime}s$. 

In Fig. 5, for (a), (b), (c), and (d), the branch is attached to the tree trunk, represented by the $W/\varphi$ propagators, through one Higgs line, the possible vertices being $HWW$, $HW\varphi$, $H\varphi W$, and $H\varphi\varphi$ three-vertices. For case (e), two Higgs lines are used to attached the branch to the trunk, where the possible vertices are $HHWW$ and $HH\varphi\varphi$. 

In this figure, the diagram of the case (f) is a symbolic representation of (b), (c), and (d). If $K$ is the momentum of the vertical Higgs line, then

\begin{equation}
\begin{aligned}
 \text{expression for} \; (f) = A \frac{1}{K^2 - m_H^2}. 
\label{aba:eq4.7}
\end{aligned}
\end{equation}

\noindent what $A$ is depends on what the branch is; two specific examples for this $A$ are
\begin{equation}
\begin{aligned}
 A = - \frac{3}{2} \frac{m_H^2}{m} 
\label{aba:eq4.8}
\end{aligned}
\end{equation}

\noindent for the branch given by (b), and 
\begin{equation}
\begin{aligned}
 A =  \frac{3}{4} \frac{m_H^2}{m^2} 
\label{aba:eq4.9}
\end{aligned}
\end{equation}

\noindent for that of (c). In both Eq.~(\ref{aba:eq4.8}) and Eq.~(\ref{aba:eq4.9}), the appearance of the factor 3 is to play an important role. This factor $A$ has momentum dependence except in these two cases. 

\begin{figure}
\begin{center}
\includegraphics[width=4in]{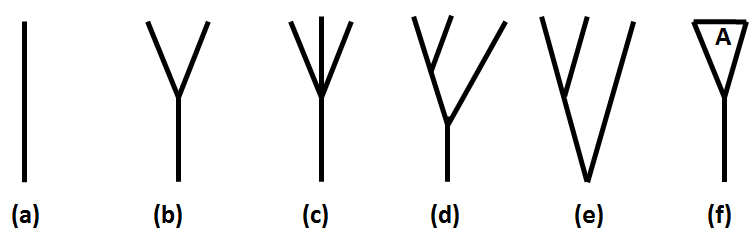}
\end{center}
\caption{Examples of the branches of a $WWnH$ tree diagram. These branches also appear in the $P$ and $Q$. Fig. 3(f) represents branches such as (b), (c), and (d), but not (a) or (e). }
\label{fig5}
\end{figure}

As seen from (\ref{aba:eq4.8}) and (\ref{aba:eq4.9}), the upper end of the vertical Higgs line in (f) is either attached to a $HHH$ three-vertex or to a $HHHH$ four-vertex. It is convenient to define $B$ and $C$ simply through
\begin{equation}
\begin{aligned}
 B = (-\frac{3}{2} \frac{m_H^2}{m})^{-1} A
\label{aba:eq4.10}
\end{aligned}
\end{equation}

\noindent in the former case, and 

\begin{equation}
\begin{aligned}
 C = (\frac{3}{4} \frac{m_H^2}{m^2})^{-1} A
\label{aba:eq4.11}
\end{aligned}
\end{equation}
\noindent in the latter case. 


\vspace{6pt}

\noindent {\bf C. Reduction of $P-Q$ --- first case}

In Sec. 4A, it has been shown by Eq.~(\ref{aba:eq4.4}) that the difference between $P^{(1)}$ and $Q^{(1)}$ is zero. Motivated by this result, the next step is to study the difference $P^{(n)} - Q^{(n)}$ recurrently in $n$. 

The simplicity of Eq.~(\ref{aba:eq4.4}) is due to the following fact: as seen from Fig.~\ref{fig3}, the segment of the horizontal line next to the one on the left is that the $W^+$. More generally, as seen from Fig.~\ref{fig4}, this segment can be that of $W^+$ or that of $\varphi^+$, the former in the case of (a) and (e), the latter in (b) and (f) for example. These two cases need to be treated separately, the $W^+$ in the present Sec.~4C and the $\varphi^+$ in the next Sec.~4D. As perhaps to the expected, the development in Sec.~4D is much more complicated than that of the present Sec.~4C. 

\begin{figure}
\begin{center}
\includegraphics[width=5in]{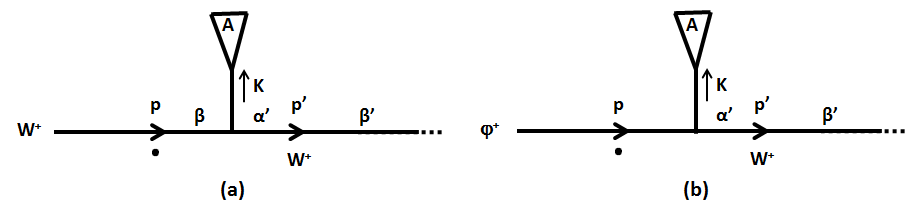}
\end{center}
\caption{Diagrams for the first case (a) $P_j^{(n)}$and (b) $Q_j^{(n)}$.}
\label{fig6}
\end{figure}

Using the notation of Fig.~\ref{fig5}(f), consider the two diagrams of Fig.~\ref{fig6}: Fig.~6(a) gives $P_j^{(n)}$ while Fig.~6(b) gives $Q_j^{(n)}$. Here the subscript $j$ indicates that these $P_j^{(n)}$ and $Q_j^{(n)}$ come from a specific diagram, a notation that has been used extensively in Fig.~\ref{fig4}. Following Sec.~3, they are given by 

\begin{equation}
\begin{aligned}
P_j^{(n)} = \frac{p^\beta}{m} A \frac{1}{K^2-m_H^2} m g_{\beta\alpha^\prime} \frac{1}{p^{\prime 2}-m^2} [-g^{\alpha^\prime \beta^\prime}  + \frac{(1-\xi)p^{\prime \alpha^\prime} p^{\prime \beta^\prime}}{p^{\prime 2}-\xi m^2}] R^{(n^\prime)}
\label{aba:eq4.12}
\end{aligned}
\end{equation}
\noindent and
\begin{equation}
\begin{aligned}
Q_j^{(n)} = A \frac{1}{K^2-m_H^2} \frac{1}{2}(p+K)_{\alpha^\prime} \frac{1}{p^{\prime 2}-m^2} [-g^{\alpha^\prime \beta^\prime}  + \frac{(1-\xi)p^{\prime \alpha^\prime} p_1^{\prime \beta^\prime}}{p^{\prime 2}-\xi m^2}] R^{(n^\prime)},
\label{aba:eq4.13}
\end{aligned}
\end{equation}

\noindent where $R^{(n^\prime)}$ denotes the factors that come from the rest of the diagram. Here $n-n^\prime$ is the number of Higgs external lines in $A$.

Since $p=p^\prime+K$, Eqs.~(\ref{aba:eq4.12}) and (\ref{aba:eq4.13}) lead immediately to 

\begin{equation}
\begin{split}
\begin{aligned}
  & P_j^{(n)} - Q_j^{(n)} \\
  &  = A \frac{1}{K^2-m_H^2} \frac{1}{2} p^\prime_{\alpha^\prime} \frac{1}{p^{\prime 2}-m^2} [-g^{\alpha^\prime \beta^\prime}  + \frac{(1-\xi)p^{\prime \alpha^\prime} p^{\prime \beta^\prime}}{p^{\prime 2}-\xi m^2}] R^{(n^\prime)} \\
  &  = A \frac{1}{K^2-m_H^2} \frac{-\xi m}{2(p^{\prime 2} - \xi m^2)} \frac{p^{\prime \beta^\prime}}{m} R^{(n^\prime)}
\label{aba:eq4.14}
\end{aligned}
\end{split}
\end{equation}

\noindent The appearance of $\frac{p^{\prime \beta^\prime}}{m}$ on the right-hand side of Eq.~(\ref{aba:eq4.12}) is important: it converts the $R^{(n^\prime)}$ into a $P$, say $P_{j^\prime}^{(n^\prime)}$: 
\begin{equation}
\begin{aligned}
P_{j^\prime}^{(n^\prime)} = \frac{p^{\prime \beta^\prime}}{m}  R^{(n^\prime)}.
\label{aba:eq4.15}
\end{aligned}
\end{equation}

\noindent Therefore
\begin{equation}
\begin{aligned}
 P_j^{(n)} - Q_j^{(n)} = \mathcal{C} A \frac{1}{K^2-m_H^2}   P_{j^\prime}^{(n^\prime)}.
\label{aba:eq4.16}
\end{aligned}
\end{equation}

\noindent where

\begin{equation}
\begin{aligned}
\mathcal{C} = \frac{-\xi m} {2 (p^{\prime 2} - \xi m^2)}.
\label{aba:eq4.17}
\end{aligned}
\end{equation}

\begin{figure}
\begin{center}
\includegraphics[width=5in]{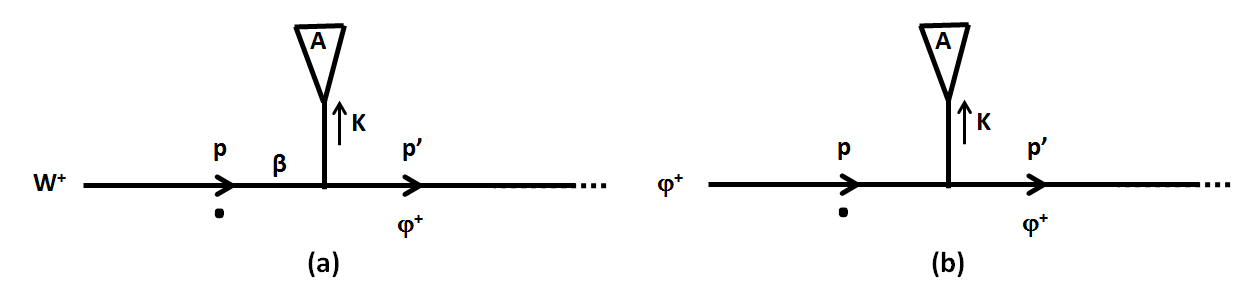}
\end{center}
\caption{Diagrams for the second case (a) $P_{j+1}^{(n)}$and (b) $Q_{j+1}^{(n)}$.}
\label{fig7}
\end{figure}

The main task at this point is to find the formula corresponding to Eq.~(\ref{aba:eq4.16}), where the $P_{j^\prime}^{(n^\prime)}$ on the right-hand side is replaced by $Q_{j^\prime}^{(n^\prime)}$ with the same $n^\prime$ and $j^\prime$. This corresponding formula must come from the diagrams of Fig.~\ref{fig6} with the $W^+$ of momentum $p^\prime$ replaced by a $\varphi^+$. These diagrams are shown in Fig.~\ref{fig7}. 

As already mentioned, this second case is much more complicated than the first case that has led to the above Eq.~(\ref{aba:eq4.16}). 

It should also be noted that everything goes through without any modification if $n-n^\prime =1 $, i.e., when $K$ is the momentum of an external Higgs particle. This is to be expected in general since this case of $n-n^\prime =1 $ is an especially simple one. 

\vspace{6pt}

\noindent {\bf D. Reduction of $P-Q$ --- second case}

Following Sec.~3C, the $P$ and $Q$ for the diagrams of Fig.~\ref{fig7} are given by 

\begin{equation}
\begin{aligned}
P_{j+1}^{(n)} = \frac{p^\beta}{m} A \frac{1}{K^2-m_H^2} \frac{1}{2} (p^\prime -K)_\beta \frac{1}{p^{\prime 2} -\xi m^2} Q_{j\prime}^{(n^\prime)}
\label{aba:eq4.18}
\end{aligned}
\end{equation}

\noindent and 
\begin{equation}
\begin{aligned}
Q_{j+1}^{(n)} = A \frac{1}{K^2-m_H^2} (-\frac{1}{2} \frac{m_H^2}{m}) \frac{1}{p^{\prime 2} -\xi m^2} Q_{j\prime}^{(n^\prime)} ,
\label{aba:eq4.19}
\end{aligned}
\end{equation}

\noindent where use has been made of the fact that the rest of the diagram gives $Q_{j^\prime}^{(n^\prime)}$.

Since 
\begin{equation}
\begin{aligned}
\frac{p^\beta}{m} \frac{1}{2} (p^\prime -K)_\beta - (-\frac{1}{2} \frac{m_H^2}{m}) = \frac{1}{2m} [p^{\prime 2} - (K^2-m^2_H)] ,
\label{aba:eq4.20}
\end{aligned}
\end{equation}

\noindent the difference $P_{j+1}^{(n)}-Q_{j+1}^{(n)}$ can be written as
\begin{equation}
\begin{aligned}
P_{j+1}^{(n)}-Q_{j+1}^{(n)} = S_0 + S_1 + S_2,
\label{aba:eq4.21}
\end{aligned}
\end{equation}

\noindent where, using the definition~(\ref{aba:eq4.17}),

\begin{equation}
\begin{aligned}
S_0 = -\mathcal{C} A \frac{1}{K^2-m_H^2} Q_{j\prime}^{(n^\prime)}
\label{aba:eq4.22}
\end{aligned}
\end{equation}

\noindent is the desired term [see Eq.~(\ref{aba:eq4.16}]), while 
\begin{equation}
\begin{aligned}
S_1 = \frac{1}{2m} A \frac{1}{K^2-m_H^2} Q_{j\prime}^{(n^\prime)}
\label{aba:eq4.23}
\end{aligned}
\end{equation}

\noindent and

\begin{equation}
\begin{aligned}
S_2 = -\frac{1}{2m} A \frac{1}{p^{\prime 2}-\xi m^2} Q_{j\prime}^{(n^\prime)}
\label{aba:eq4.24}
\end{aligned}
\end{equation}

\noindent need to be combined with the contributions from additional diagrams. In the notations $S_0$, $S_1$, and $S_2$, the indices $j^\prime+1$ and $n^\prime$ have been suppressed. The suppression of these and other indices will also be applied to a number of additional quantities in this Sec.~3D. 

\begin{figure}
\begin{center}
\includegraphics[width=5in]{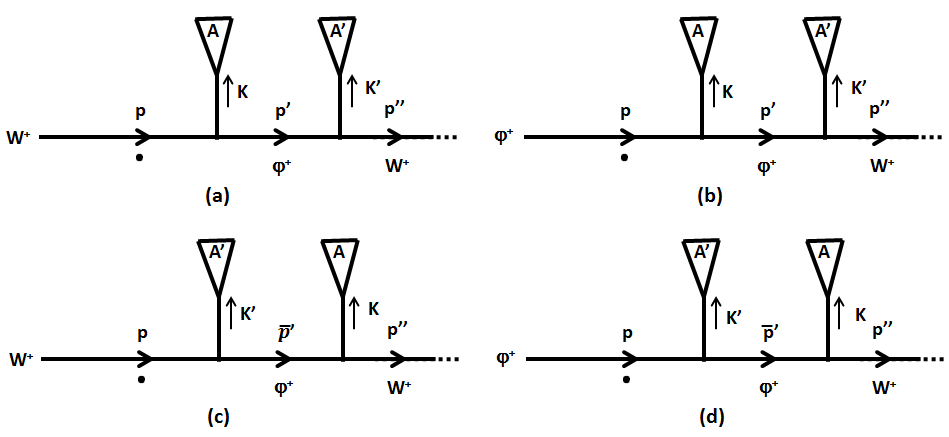}
\end{center}
\caption{Diagrams for some of the contributions to $P_{j+1}^{(n)}$ and $Q_{j+1}^{(n)}$ (a) $P_a$ and (b) $Q_a$. The diagrams (c) $P_b$ and (d) $Q_b$ are obtained from (a) and (b) by exchanging two branches of the tree.}
\label{fig8}
\end{figure}

As seen from Figs.~\ref{fig6} and \ref{fig7}, the momentum of the propagator to the left has been called $p$ and that next-to-left called $p^\prime$. In order to study additional contributions similar to $S_1$ and $S_2$, it is necessary to introduce a further momentum $p^{\prime\prime}$. This leads to the diagrams of Figs.~\ref{fig8}(a) and \ref{fig8}(b); when the two branches in these diagrams are exchanged, they then lead to those Figs.~\ref{fig8}(c) and \ref{fig8}(d).

As noted in the caption of Fig.~\ref{fig8}, $P_a$ and $Q_a$ are parts of $P_{j+1}^{(n)}$ and $Q_{j+1}^{(n)}$ respectively, the indices $j+1$ and $n$ having been suppressed. It therefore follows form Eqs.~(\ref{aba:eq4.21}) and (\ref{aba:eq4.23}) that 

\begin{equation}
\begin{aligned}
P_a - Q_a = S_{0a} + S_{1a} + S_{2a}, 
\label{aba:eq4.25}
\end{aligned}
\end{equation}
\noindent where

\begin{equation}
\begin{aligned}
S_{1a} = \frac{1}{2m} A A^\prime \frac{1}{K^2-m_H^2} \frac{1}{K^{\prime 2}-m_H^2} \frac{1}{2} (p^\prime + K^\prime)_{\alpha^{\prime\prime}} R^{(n^{\prime\prime})}.
\label{aba:eq4.26}
\end{aligned}
\end{equation}

\noindent Here $ R^{(n^{\prime\prime})}$, similar to the $ R^{(n^{\prime})}$ of Eqs.~(\ref{aba:eq4.12})-(\ref{aba:eq4.15}), are the factors that come from the rest of the diagram. When the two branches of the tree are exchanged, leading to the diagrams of Figs.~\ref{fig8}(c) and \ref{fig8}(d) for $P_b$ and $Q_b$, the corresponding expressions are 

\begin{equation}
\begin{aligned}
P_b - Q_b = S_{0b} + S_{1b} + S_{2b}, 
\label{aba:eq4.27}
\end{aligned}
\end{equation}

\noindent where

\begin{equation}
\begin{aligned}
S_{1b} = \frac{1}{2m} A A^\prime \frac{1}{K^2-m_H^2} \frac{1}{K^{\prime 2}-m_H^2} \frac{1}{2} (\bar{p}^\prime + K)_{\alpha^{\prime\prime}} R^{(n^{\prime\prime})}.
\label{aba:eq4.28}
\end{aligned}
\end{equation}

\noindent Note that the same $R^{(n^{\prime\prime})}$ appears in Eqs.~(\ref{aba:eq4.26}) and~(\ref{aba:eq4.28}). Furthermore, as seen from Fig.~\ref{fig8}, the momenta are related by 

\begin{center}
$p^\prime = p - K;$
\end{center}

\noindent and

\begin{equation}     
\begin{split}
\begin{aligned}
&\;\;\, \bar{p}^\prime = p - K^\prime;
\label{aba:eq4.29}
\end{aligned}
\end{split}
\end{equation}

\noindent thus

\begin{equation}
\begin{aligned}
(p^\prime + K^\prime) + (\bar{p}^\prime + K) = 2p. 
\label{aba:eq4.30}
\end{aligned}
\end{equation}

\noindent It therefore follows from Eqs.~(\ref{aba:eq4.26}) and~(\ref{aba:eq4.28}) that

\begin{equation}
\begin{aligned}
S_{1a} + S_{1b} = \frac{1}{2m} A A^\prime \frac{1}{K^2-m_H^2} \frac{1}{K^{\prime 2}-m_H^2} p_{\alpha^{\prime\prime}} R^{(n^{\prime\prime})}.
\label{aba:eq4.31}
\end{aligned}
\end{equation}

\begin{figure}
\begin{center}
\includegraphics[width=3.5in]{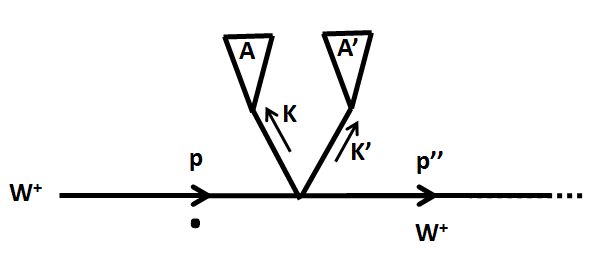}
\end{center}
\caption{Diagrams for another contribution, called $P_c$, to $P_{j+1}^{(n)}$.}
\label{fig9}
\end{figure}

This expression is very nice: it means that 
\begin{equation}
\begin{aligned}
S_{1a} + S_{1b} + P_c = 0
\label{aba:eq4.32}
\end{aligned}
\end{equation}

\noindent where $P_c$ is the contribution to $P_{j+1}^{(n)}$ coming from the diagram of Fig.~\ref{fig9}. The corresponding $Q_c$ is zero because there is no $W\varphi HH$ four-vertex. This takes care of all the $S_1^\prime s$ from Eqs.~(\ref{aba:eq4.21}) and~(\ref{aba:eq4.23}).

\begin{figure}
\begin{center}
\includegraphics[width=5in]{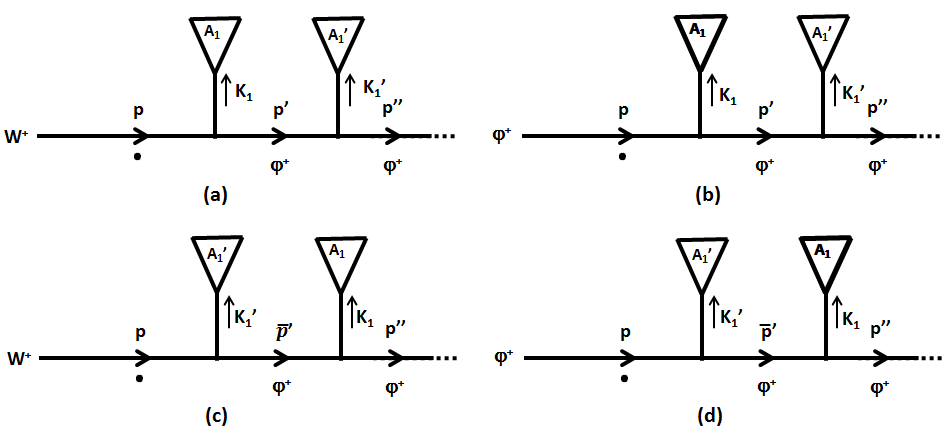}
\end{center}
\caption{Diagrams for some further contributions to $P_{j+1}^{(n)}$ and $Q_{j+1}^{(n)}$ (a) $P_d$ and (b) $Q_d$. The diagrams (c) $P_e$ and (d) $Q_e$ are obtained from (a) and (b) by exchanging two branches with $A$, and $A_1^\prime$.}
\label{fig10}
\end{figure}

It remains to study the $S_2$ as given by Eqs.~(\ref{aba:eq4.21}) and~(\ref{aba:eq4.24}). As seen from Fig.~\ref{fig8}, the diagrams to be studied are the two shown in Fig.~\ref{fig10}. The diagrams in this Fig.~\ref{fig10} differ from those of Fig.~\ref{fig8} in that it is a $\varphi^+$ propagator with momentum $p^{\prime\prime}$ instead of a $W^+$ propagator. For Figs.~\ref{fig10}(a) and \ref{fig10}(b), the $P_d$ and $Q_d$ are given by 
\begin{equation}
\begin{aligned}
P_d = \frac{p^\beta}{m} A_1 \frac{1}{K_1^2 - m_H^2} \frac{1}{2} (p^\prime -K_1)_\beta \frac{1}{p^{\prime 2} - \xi m^2}  A_1^\prime   \frac{1}{K_1^{\prime 2} - \xi m^2}  R_d^{(n^\prime)}
\label{aba:eq4.33}
\end{aligned}
\end{equation}

\noindent and

\begin{equation}
\begin{aligned}
Q_d = A_1 \frac{1}{K_1^2 - m_H^2} (-\frac{1}{2} \frac{m_H^2}{m})  \frac{1}{p^{\prime 2} - \xi m^2}  A_1^\prime   \frac{1}{K_1^{\prime 2} - m_H^2}  R_d^{(n^\prime)}.
\label{aba:eq4.34}
\end{aligned}
\end{equation}

\noindent It follows from Eq.~(\ref{aba:eq4.20}) that the difference $P_d - Q_d$ can be written as the sum of three terms
\begin{equation}
\begin{aligned}
P_d - Q_d = T_0 + T_1 + T_2
\label{aba:eq4.35}
\end{aligned}
\end{equation}
\noindent where 

\begin{equation}
\begin{aligned}
T_0 = -\mathcal{C} A_1 \frac{1}{K_1^2 - m_H^2}  A_1^\prime   \frac{1}{K_1^{\prime 2} - m_H^2}   R_d^{(n^\prime)},
\label{aba:eq4.36}
\end{aligned}
\end{equation}

\begin{equation}
\begin{aligned}
T_1 = \frac{1}{2m} A_1 \frac{1}{K_1^2 - m_H^2}  A_1^\prime   \frac{1}{K_1^{\prime 2} - m_H^2}   R_d^{(n^\prime)},
\label{aba:eq4.37}
\end{aligned}
\end{equation}

\noindent and
\begin{equation}
\begin{aligned}
T_2 = -\frac{1}{2m} A_1 \frac{1}{p^{\prime 2} - \xi m^2}  A_1^\prime   \frac{1}{K_1^{\prime 2} - m_H^2}   R_d^{(n^\prime)}.
\label{aba:eq4.38}
\end{aligned}
\end{equation}

\noindent A comparison with  Eqs.~(\ref{aba:eq4.21}) -~(\ref{aba:eq4.24}) shows that these $T's$ are special case of the $S's$ Since it is $S_2$ being studied, this $T_2$ can be treated accordingly. It therefore remains to consider the $T_1$ of Eq.~(\ref{aba:eq4.37}). 

Figs.~\ref{fig8}(c) and \ref{fig8}(d) are obtained from Figs.~\ref{fig8}(a) and \ref{fig8}(b) by exchanging two branches of the tree; entirely similarly, Figs.~\ref{fig10}(c) and \ref{fig10}(d) are obtained from Figs.~\ref{fig10}(a) and \ref{fig10}(b) by the same exchange. In this way, Eq.~(\ref{aba:eq4.35}) leads to 

\begin{equation}
\begin{aligned}
P_e - Q_e = \hat{T}_0 + \hat{T}_1 + \hat{T}_2,
\label{aba:eq4.39}
\end{aligned}
\end{equation}

\noindent where the $\hat{}$ denotes the exchange of the two branches, a generalization of the notation already used in Eqs.~(\ref{aba:eq4.5}) and~(\ref{aba:eq4.6}). As seen from Eq.~(\ref{aba:eq4.37}), however, $T_1$ is invariant under this exchange; therefore

\begin{equation}
\begin{aligned}
\hat{T}_1 = T_1.
\label{aba:eq4.40}
\end{aligned}
\end{equation}

\begin{figure}
\begin{center}
\includegraphics[width=3.5in]{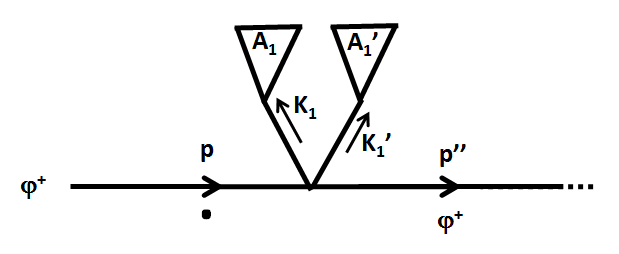}
\end{center}
\caption{Diagrams for another contribution, called $Q_f$, to $Q_{j+1}^{(n)}$.}
\label{fig11}
\end{figure}

\noindent Moreover, there is a third similar contribution from the diagram of Fig.~\ref{fig11}. In this case, by the Feynman rule of Fig.~\ref{fig2}, this $Q_f$ is given by 

\begin{equation}
\begin{aligned}
Q_f = A_1 \frac{1}{K_1^2 - m_H^2} A_1^\prime \frac{1}{K_1^{\prime 2} - m_H^2} (\frac{1}{4} \frac{m_H^2}{m^2}) (-\frac{1}{2} \frac{m_H^2}{m})^{-1}  R_d^{(n^\prime)}.
\label{aba:eq4.41}
\end{aligned}
\end{equation}

\noindent The corresponding $P_f$ is zero again because there is no $W\varphi HH$ four-vector. A comparison of this expression with Eq.~(\ref{aba:eq4.37}{}) shows immediately that 

\begin{equation}
\begin{aligned}
Q_f = - T_1. 
\label{aba:eq4.42}
\end{aligned}
\end{equation}

\noindent Together with Eq.~(\ref{aba:eq4.40}), the result is 

\begin{equation}
\begin{aligned}
T_1 + \hat{T}_1 - Q_f = 3 T_1, 
\label{aba:eq4.43}
\end{aligned}
\end{equation}

\noindent where the minus sign before $Q_f$ comes from $P_f - Q_f $ = $-Q_f$. It remains to combine this result with the $S_2$ given by Eq.~(\ref{aba:eq4.24}). There are two distinct situations to be studied, as decided at the end of Sec.~4B. 

As seen from Fig.~\ref{fig11} and Eq.~(\ref{aba:eq4.41}) for example, two Higgs internal lines, with momenta $K_1$ and $K_1^\prime$, are attached to the $W^+/\varphi^+$ line. Therefore, for the present purpose, Eq.~(\ref{aba:eq4.10}) is to be used, not Eq.~(\ref{aba:eq4.11}). Furthermore, the identifications are

\begin{equation}
\begin{aligned}
B = A_1 \frac{1}{K_1^2 - m_H^2} A_1^\prime \frac{1}{K_1^{\prime 2} - m_H^2}
\label{aba:eq4.44}
\end{aligned}
\end{equation}

\noindent and 

\begin{equation}
\begin{aligned}
\frac{1}{p^{\prime 2} - \xi m^2} Q_{j^\prime}^{(n^\prime)} = (-\frac{1}{2} \frac{m_H^2}{m})^{-1}  R_d^{(n^\prime)}.
\label{aba:eq4.45}
\end{aligned}
\end{equation}

\noindent With these  Eqs.~(\ref{aba:eq4.44}) and~(\ref{aba:eq4.45}), the $S_2$ of Eq.~(\ref{aba:eq4.24}) takes the form 

\begin{equation}
\begin{aligned}
S_2 = - \frac{3}{2m} A_1 \frac{1}{K_1^2 - m_H^2} A_1^\prime \frac{1}{K_1^{\prime 2} - m_H^2} R_d^{(n^\prime)}, 
\label{aba:eq4.46}
\end{aligned}
\end{equation}

\noindent which leads to the cancellation, because of  Eqs.~(\ref{aba:eq4.37}) and~(\ref{aba:eq4.43}), 

\begin{equation}
\begin{aligned}
S_2 + T_1 + \hat{T}_1 - Q_f = 0.
\label{aba:eq4.47}
\end{aligned}
\end{equation}

In the second situation Eq.~(\ref{aba:eq4.11}) is to be used instead of Eq.~(\ref{aba:eq4.10}), i.e., where the vertical line of Fig.~\ref{fig5}(f) is attached to a $HHHH$ four-vertex instead of a $HHH$ three-vertex. Therefore $C$ consists of three factors of the form

\begin{equation}
\begin{aligned}
C = A_1 \frac{1}{K_1^2 - m_H^2} A_1^\prime \frac{1}{K_1^{\prime 2} - m_H^2} A_1^{\prime\prime} \frac{1}{K_1^{\prime\prime 2} -m_H^2}
\label{aba:eq4.48}
\end{aligned}
\end{equation}

\noindent  and
\begin{equation}
\begin{aligned}
S_2 = - \frac{1}{2m} A_1 \frac{1}{K_1^2 - m_H^2} A_1^\prime \frac{1}{K_1^{\prime 2} - m_H^2} A_1^{\prime\prime} \frac{1}{K_1^{\prime\prime} -m_H^2} (\frac{3}{4} \frac{m_H^2}{m^2}) \frac{1}{p^{\prime 2} - \xi m^2} Q_{j^\prime}^{(n^\prime)}. 
\label{aba:eq4.49}
\end{aligned}
\end{equation}

\begin{figure}
\begin{center}
\includegraphics[width=5in]{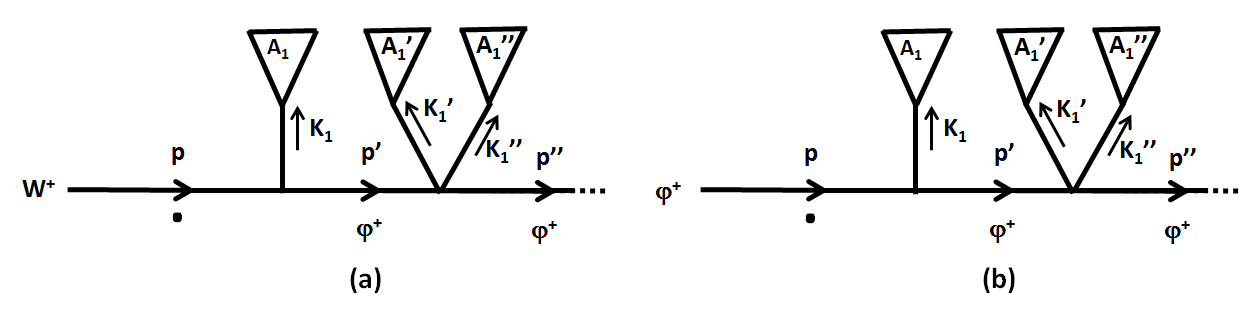}
\end{center}
\caption{Diagrams that give contributions similar to the right-hand side of Eq.~(\ref{aba:eq4.49}) (a) $P_g$ and (b) $Q_g$. Cyclic permutation of $(A_1, K_1), (A_1^{\prime}, K_1^{\prime}), (A_1^{\prime\prime}, K_1^{\prime\prime})$ gives additional contributions designated by $P_{g^\prime}, Q_{g^\prime},  P_{g^{\prime\prime}},$ and $Q_{g^{\prime\prime}}$.}
\label{fig12}
\end{figure}

\noindent Because of Figs.~\ref{fig9} and \ref{fig11} together with Eqs.~(\ref{aba:eq4.32}) and~(\ref{aba:eq4.42}), the only diagrams that can give contributions similar to right-hand side of Eq.~(\ref{aba:eq4.49}) are that of Fig.~\ref{fig12} together with the two obtained by the cyclic permutation  of $(A_1, K_1), (A_1^{\prime}, K_1^{\prime}), (A_1^{\prime\prime}, K_1^{\prime\prime})$. Let these three contributions be called $P_g, Q_g; P_{g^\prime}, Q_{g^\prime}; $ and $P_{g^{\prime\prime}}, Q_{g^{\prime\prime}}$, then it follow from Eqs.~(\ref{aba:eq4.35}) and~(\ref{aba:eq4.37}) that 

\begin{equation}
\begin{aligned}
P_g - Q_g = T_{0g} + T_{1g} + T_{2g},
\label{aba:eq4.50}
\end{aligned}
\end{equation}

\noindent where

\begin{equation}
\begin{aligned}
T_{1g} = \frac{1}{2m} A_1 \frac{1}{K_1^2 - m_H^2} A_1^\prime \frac{1}{K_1^{\prime 2} - m_H^2} A_1^{\prime\prime} \frac{1}{K_1^{\prime\prime 2} -m_H^2}  R_g^{(n^\prime)}, 
\label{aba:eq4.51}
\end{aligned}
\end{equation}

\noindent while $T_{0g}$ and $T_{2g}$ can be treated as $T_0$ and $T_2$ without causing any difficulty. Since, as seen from Fig.~\ref{fig12}, $ R_g^{(n^\prime)}$ are the factors that come from the rest of the diagram, it does not change when the three branches are permuted, implying that

\begin{equation}
\begin{aligned}
T_{1g} = T_{1g^\prime} = T_{1g^{\prime\prime}}.
\label{aba:eq4.58}
\end{aligned}
\end{equation}

\noindent In particular, for the present case of the $HHHH$ four-vertex, the analog of Eq.~(\ref{aba:eq4.43}) is 

\begin{equation}
\begin{aligned}
T_{1g} + T_{1g^\prime} + T_{1g^{\prime\prime}} = 3 T_{1g}.
\label{aba:eq4.53}
\end{aligned}
\end{equation}

\noindent This is due to the fact that there is no term corresponding to the $Q_f$ coming from the diagram of Fig.~\ref{fig11} because there is no five-vertex in the standard model. 

Together with Eq.~(\ref{aba:eq4.48}), the identification 

\begin{equation}
\begin{aligned}
\frac{1}{p^{\prime 2} - \xi m^2} Q_{j^\prime}^{(n^\prime)} = (\frac{3}{4} \frac{m_H^2}{m^2})^{-1}  R_g^{(n^\prime)}
\label{aba:eq4.54}
\end{aligned}
\end{equation}

\noindent for the present case instead of  Eq.~(\ref{aba:eq4.45}) then gives the cancellation

\begin{equation}
\begin{aligned}
S_2 + T_{1g} + T_{1g^\prime} + T_{1g^{\prime\prime}} = 0.
\label{aba:eq4.55}
\end{aligned}
\end{equation}

\noindent In obtaining the cancellations of both Eq.~(\ref{aba:eq4.47}) and Eq.~(\ref{aba:eq4.55}), a factor of 3 plays a prominent role, a point already noted after Eq.~(\ref{aba:eq4.9})

This completes the treatment of the diagrams shown in Fig.~\ref{fig7}.

\vspace{6pt}
\noindent {\bf E. Cancellations and recurrence relations}

In the first case treated in Sec.~4C, the result is given by  Eq.~(\ref{aba:eq4.16}). In this equation, the subscript $j$ refers to a diagram, for either $P$ or $Q$, consisting of a branch on the left-hand side and the subscript $j^\prime$ refers to the rest of the diagram. If this branch is designated as $br$, then

\begin{equation}
\begin{aligned}
j = (br, j^\prime).
\label{aba:eq4.56}
\end{aligned}
\end{equation}

\noindent The right-hand side of this Eq.~(\ref{aba:eq4.16}) is a product of three factors: a $\mathcal{C}$ defined by Eq.~(\ref{aba:eq4.17}), a factor that depends on the branch $br$, and then a factor that depends on the rest of the diagram. The superscripts refer to the number of external Higgs lines, where $n-n^\prime$ is the number of these lines for branch $br$. Note that while $P-Q$ is on the left-hand side of this equation, only $P$ appears on the right. 

In contrast, the second case treated in Sec.~4D is much more complicated. As seen from Eq.~(\ref{aba:eq4.21}) there, the right-hand side consists not only the desired term $S_0$, but also two additional terms $S_1$, and $S_2$. The cancellation of $S_1$ terms is expressed by Eq.~(\ref{aba:eq4.32}), where the appearance of the term $P_c$ justifies the factor $\frac{1}{K^2-m_H^2}$ in Eq.~(\ref{aba:eq4.16}) for the first case. 

The corresponding cancellation of the $S_2$ terms of Eq.~(\ref{aba:eq4.21}) is much less straightforward. The basic formulas are given by Eqs.~(\ref{aba:eq4.47}) and~(\ref{aba:eq4.53}). However, in order to get the cancellations of the $S_2$ terms, these Eqs.~(\ref{aba:eq4.47}) and~(\ref{aba:eq4.53}) need to be applied repeatedly. After the cancellation of both the $S_1$ and the $S_2$ terms, only $S_0$ remains and the resulting expression can be combined with Eq.~(\ref{aba:eq4.16}) of Sec.~4C. For the purpose of summing over $j^\prime$, the following notation is convenient

\begin{equation*}
\begin{aligned}
P_{br}^{(n)} = \sum_{j^\prime} P_j^{(n)}
\end{aligned}
\end{equation*}

\noindent and 
\begin{equation}
\begin{aligned}
Q_{br}^{(n)} = \sum_{j^\prime} Q_j^{(n)},
\label{aba:eq4.58}
\end{aligned}
\end{equation}

\noindent where the notation (\ref{aba:eq4.56}) has been used. It then follows from the result of Secs.~4C and 4D that 

\begin{equation}
\begin{aligned}
P_{br}^{(n)} - Q_{br}^{(n)} = \mathcal{C} A \frac{1}{K^2-m_H^2} (P^{(n^\prime)} - Q^{(n^\prime)}),
\label{aba:eq4.58}
\end{aligned}
\end{equation}

\noindent where $A$ and $K$ depend on the subscript $br$. Since 
\begin{equation*}
\begin{aligned}
 P^{(n)} = \sum_{br} P_{br}^{(n)} 
\end{aligned}
\end{equation*}
\begin{equation}
\begin{aligned}
 Q^{(n)} = \sum_{br} Q_{br}^{(n)}, 
\label{aba:eq4.59}
\end{aligned}
\end{equation}

\noindent Eq.~(\ref{aba:eq4.58}) gives

\begin{equation}
\begin{aligned}
P^{(n)} - Q^{(n)} = \mathcal{C} \sum_{br} A_{br} \frac{1}{K^2_{br}-m_H^2} (P^{(n^\prime)} - Q^{(n^\prime)}),
\label{aba:eq4.60}
\end{aligned}
\end{equation}

\noindent where, for clarity, the subscript $br$ has been added back to $A$ and $K$.

Because of Eq.~(\ref{aba:eq4.4}) and the fact that there is at least one Higgs external line in the branch $br$ (i.e., $n-n^\prime \geq 1$), mathematical induction can be applied to Eq.~(\ref{aba:eq4.60}) to give the desired result

\begin{equation}
\begin{aligned}
P^{(n)} - Q^{(n)} = 0
\label{aba:eq4.61}
\end{aligned}
\end{equation}

\noindent for all $n$. 

\vspace{6pt}
\noindent {\bf F. Gauge invariance}

Eq.~(\ref{aba:eq4.61}) implies immediately that gauge invariance hold for the case under consideration: matrix elements on the tree level with the two external $W$ bosons and any number of external Higgs bosons. The argument is as follows. 

Eq.~(\ref{aba:eq4.61}) implies that 

\begin{equation}
\begin{aligned}
P^{\prime(n)} - Q^{\prime(n)} = 0
\label{aba:eq4.62}
\end{aligned}
\end{equation}

\noindent for all $n$. Let the number of external Higgs bosons be called $N$, then the derivative with respect to $\xi$ of this tree matrix element is a sum of terms proportional to $P^{(n)} P^{\prime(n^\prime)} - Q^{(n)} Q^{\prime(n^\prime)}$ with $n+n^\prime = N$. But 
\begin{equation}
\begin{aligned}
P^{(n)} P^{\prime(n^\prime)} - Q^{(n)} Q^{(n^\prime)}= 0
\label{aba:eq4.63}
\end{aligned}
\end{equation}
\noindent as a consequence of Eqs.~(\ref{aba:eq4.61}) and~(\ref{aba:eq4.62}). Therefore, this derivative with respect to $\xi$ is zero, and hence the matrix-element itself is a constant. Since this matrix element is a rational function of $\xi$, this being a constant implies that gauge invariance holds in this case; see the detailed discussion in Sec.~2. 

\section{Conclusion and discussions}
In Ref.~\cite{ttwu201}, it is shown that formal conclusion are not necessarily valid in every case. Rather, such formal conclusions need to be checked by explicit calculations. In that reference, two examples are given where gauge invariance supported by formal arguments turns out to be violated. These two example are for the decay processes~(\ref{aba:eq1.1}) and~(\ref{aba:eq1.2}). 

In view of these two counter-examples, it is desirable to find significant classes of non-trivial examples where formal argument and explicit calculation do lead to the same conclusion on the validity of gauge investigation. 

The class of diagrams studied consists of tree diagrams with two external $W$ bosons and any number of external Higgs bosons. For this large class of matrix elements, the explicit calculation, carried out through mathematical induction on the number of external Higgs bosons, confirms the formal result that gauge invariance does hold. This is the first time such a verification has been accomplished for a large class of cases. These cases are non-trivial because the expressions for the matrix elements are quite different in the $R_\xi$ gauge and in the unitary gauge. 

As expected for such a large class of complicated diagrams as seen from Fig.~\ref{fig1} for example, the derivation of the results through explicit calculation is quite lengthy. Nevertheless, the present derivation by induction involves much less work than a brute force computation even for a moderate value of the number of external Higgs bosons. Moreover, the present result gives strong indication that there is no gauge non-invariance for all tree diagrams. 

Combined with the result of Ref.~\cite{ttwu201} and also Ref.~\cite{rgastman}, the situation is as follows: there is no violation of gauge invariance for tree diagrams but there is for some one-loop diagrams. Thus the present paper may be considered also as proposing the first step in the development of a systematic method to calculate the difference between various gauges. This method, which has been applied only to tree diagrams in the present paper, needs to be generalized first to one-loop diagrams and then to multi-loop diagrams. 

Let us conclude this paper by repeating the final sentence of Ref.~\cite{ttwu201}: 

{\bf Yang-Mills non-Abelian gauge theory~\cite{CNYRL} in general and the standard model in particular are much more subtle than what has been generally realized.}  

\section{Acknowledgements}

We are most grateful to Professor Chen Ning Yang for numerous discussion through many years. For helpful discussions, we also thank  Raymond Gastmans, Andr\'{e} Martin, John Myers, Jacques Soffer, and Ivan Todorov. We are  indebted to the hospitality at CERN, where part of this work has been carried out.



\begin{thebibliography}{00}

\bibitem{ttwu201} T. T. Wu, S. L. Wu, {\it Nucl. Phys. B}   914  (2017) 421.

\bibitem{sglashow} S. L. Glashow, {\it Nucl. Phys.} 22 (1961) 579;  \\S. Weinberg, {\it Phys. Rev. Lett. } 19 (1967) 1264; \\A. Salam, in N. Svartholm (Ed.) Proc. 8th. Nobel Symp. Stockholm, 1968, Almgvist, 1968, p. 367. 

\bibitem{FERB} F. Englert, R. Brout, {\it Phys. Rev. Lett.} 13 (1964) 321; \\P. W. Higgs, {\it Phys. Lett.} 12, (1964) 132; \\G. S. Guralnik, C. R. Hagen, T. W. B. Kibble, {\it Phys. Rev. Lett.} 13 (1964) 585.

\bibitem{atlasdisc} ATLAS Collaboration, {\it Phys. Lett. B} 716 (2012) 1; \\CMS Collaboration, {\it Phys. Lett. B} 716 (2012) 30.

\bibitem{lfadv} L. Faddeev, V. Popov, {\it Phys. Lett. B} 25 (1967) 29.

\bibitem{schwinger} J. Schwinger, {\it Phys. Rev.} 91 (1953) 713, see especially p. 720 and p. 723; \\J. Schwinger, {\it Phys. Rev.} 94 (1954) 1362, see especially Eq. (54) on p. 1366 and the discussion after Eq. (208) on p. 1376; \\G. L\"{u}ders, Kgl. Danske Videnskab. Selskab, Mat.-fys. Medd. 28 (1954) No. 5. 

\bibitem{rgastman} R. Gastmans, S. L. Wu, T. T. Wu, CERN preprint CERN-PH-TH/2011-200, arXiv:1108.5322 2011; \\R. Gastmans, S. L. Wu, T. T. Wu, CERN preprint CERN-PH-TH/2011-201, arXiv:1108.5872 2011;\\  R. Gastmans, S. L. Wu, T. T. Wu,   {\it Int. J. Mod. Phys. A} 30 (2015) 1550200; \\T. T. Wu, S. L. Wu, {\it Int. J. Mod. Phys. A}  31 (2016) 1650028.

\bibitem{CNYRL} C. N. Yang, R. L. Mills, {\it Phys. Rev.} 96 (1954) 191.


\end{thebibliography}
\end{document}